\documentclass[conference]{IEEEtran}
\usepackage{url}
\usepackage{latexsym}

 \usepackage{caption}
\usepackage{algorithm}

\usepackage{graphicx}
\usepackage[english]{babel}
\usepackage{epstopdf}
\usepackage{pfnote}
\usepackage{multirow}
\usepackage{xcolor}
\usepackage{pgfplots}
\usepackage{tikz}
\usepackage{algpseudocode}
\usepackage{subcaption}
            
\usepackage[justification=centering]{caption}
\usepackage{float}
\usepackage{wrapfig}
\usetikzlibrary{patterns}

\usepackage{dblfnote}
\DFNalwaysdouble 

\usepackage[nopar]{lipsum}

\newcommand{\nop}[1]{}

\usepackage{mdwlist}

\newif\ifquoteopen
\catcode`\"=\active 
\DeclareRobustCommand*{"}{%
   \ifquoteopen
     \quoteopenfalse 
   \else
     \quoteopentrue 
   \fi
}

\newenvironment{customlegend}[1][]{%
    \begingroup
    \csname pgfplots@init@cleared@structures\endcsname
    \pgfplotsset{#1}%
}{%
    \csname pgfplots@createlegend\endcsname
    \endgroup
}%

\def\addlegendimage{\csname pgfplots@addlegendimage\endcsname}

\ifCLASSINFOpdf
\else
\fi
\begin{document}
%
\title{FLATM: A Fuzzy Logic Approach Topic Model for Medical Documents}

\author{\IEEEauthorblockN{Amir Karami, Aryya Gangopadhyay, Bin Zhou}
\IEEEauthorblockA{Information Systems Department\\
University of Maryland Baltimore County\\
Baltimore, Maryland 21250\\
Email: amir.karami, gangopad, bzhou@umbc.edu}
\and
\IEEEauthorblockN{Hadi Kharrazi}
\IEEEauthorblockA{Bloomberg School of Public Health \\
Johns Hopkins University \\
Baltimore, Maryland 21205\\
Email: kharrazi@jhu.edu}}


%


\maketitle

\begin{abstract}
One of the challenges for text analysis in medical domains is analyzing large-scale medical documents. As a consequence, finding relevant documents has become more difficult. One of the popular methods to retrieve information based on discovering the themes in the documents is topic modeling. The themes in the documents help to retrieve documents on the same topic with and without a query. In this paper, we present a novel approach to topic modeling using fuzzy clustering. To evaluate our model, we experiment with two text datasets of medical documents. The evaluation metrics carried out through document classification and document modeling show that our model produces better performance than LDA, indicating that fuzzy set theory can improve the performance of topic models in medical domains.
\end{abstract}


%
\IEEEpeerreviewmaketitle

\section{Introduction}

In the past several years, the medical data have been growing explosively. For example, PubMed\footnote{http://www.ncbi.nlm.nih.gov/pubmed} is one of the biggest databases for medical research articles. The statistics of PubMed show that the number of papers published in PubMed was increased from 112,177 in 1960 to 2,019,238 in 2013 and the growth rate of publication between 2010 and 2013 is more than 200\%. As another example, hospital documents are one major type of medical data. Based on the statistics of U.S. Department of Health and Human Services\footnote{http://www.icpsr.umich.edu/icpsrweb/ICPSR/studies/6222/version/1}, the annual average number of discharges between 2007 and 2010 is  around 35 million records. Analyzing such large-scale medical data is of great importance to enhance health care for millions of people. As reported in \cite{kohn2000err}, more than 44,000 patients died in the hospital as a result of medical errors. In addition, the healthcare industry could save \$450 billion a year using advanced data analytical approaches\footnote{http://www.mckinsey.com/insights/health\_systems\_and\_services/the\_big-data\_revolution\_in\_us\_health\_care}.

However, as the majority of medical data are in unstructured free-text format, there is a big challenge to develop methods to analyze large-scale unstructured medical data. Recently, various text mining techniques have been introduced into the medical domain. One fundamental objective of those techniques is to process the unstructured medical data into a proper format for better utilization to recognize explicit facts. Due to the natural probabilistic reasoning of unstructured text data, \emph{topic model} such as Latent Dirichlet Allocation (LDA) \cite{blei2003latent} has attracted much attention for analyzing medical data. Topic model is one type of statistical models for discovering the latent ``topics'' that occur in a document collection. It is able to provide a representation of free-text documents in terms of latent features discovered from the collection to generalize an algorithm to unseen documents (Figure 1). Several recent research studies have applied topic models on medical data for different purposes, such as medical document categorization \cite{sarioglu2012clinical, sarioglu2013topic}, medical document retrieval \cite{zeng2012synonym, huang2014discovery, as012009predicting}, medical document analysis \cite{arnold2010clinical,howes2013using}, etc.

\begin{figure}[H]
  \caption{The Intuition Behind LDA}
  \centering
    \includegraphics[width=8cm]{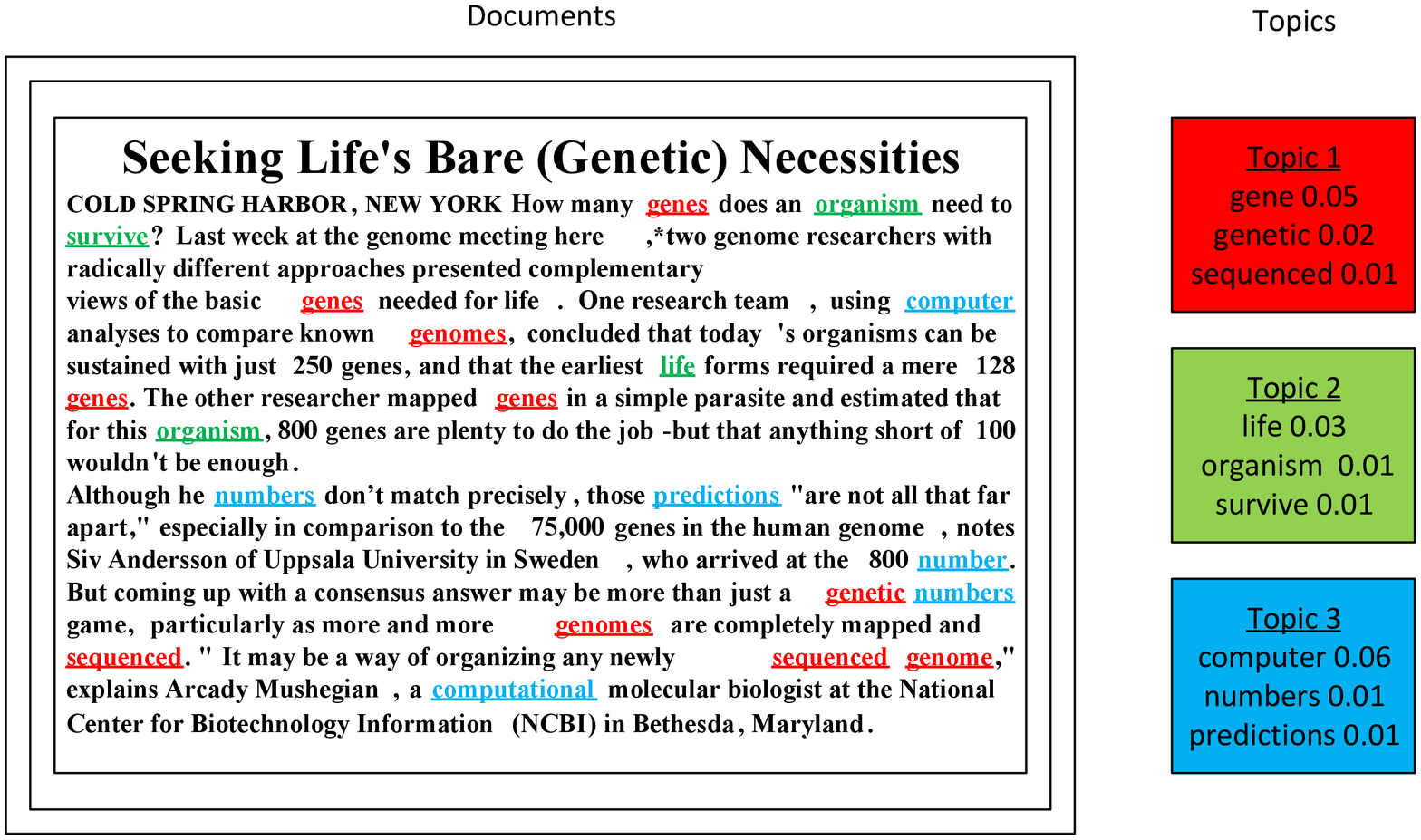}
\end{figure}

Despite the usefulness of topic models for medical data analysis \cite{arnold2010clinical,bisgin2012investigating}, existing topic models such as LDA still suffer from several critical issues. One issue of those existing topic models is their computational complexity. Almost all uses of topic models require probabilistic inference, which is arguably hard to achieve without approximate inference algorithms such as Gibbs sampling. Another issue of those existing topic models is their expressive power of representing medical documents.

The performance of various tasks such as document classification and modeling using topic models is still not satisfactory. In this paper, we propose to model medical documents using fuzzy set theory. Fuzzy set theory models membership of objects using a possibility distribution. Most of the studies using fuzzy set theory in the medical domain are related to image processing \cite{cui2013global,beevi2012robust}. A few work have been done in medical text mining using fuzzy clustering \cite{ben2012data,fenza2012hybrid}. 

\newpage
The main difference between our method and other document fuzzy clustering methods such as \cite{singh2011document} is that our method uses fuzzy clustering and word weighting as a pre-processing step for feature transformation before implementing any supervised or unsupervised algorithms; however, other methods use fuzzy clustering as a final step to cluster or classify the documents. 

To the best of our knowledge, this is the first study in the medical domain that has been done to use fuzzy set theory to express semantic properties of words and documents in terms of topics. Ideally, if we model words in the documents as objects and a group of relevant words as a latent ``topic,'' the fuzzy set theory provides a natural probabilistic view of free-text documents. Compared with existing topic models such as LDA, the fuzzy set theory is computationally efficient to achieve. We develop several efficient strategies to model medical documents using fuzzy set theory.

Regarding the expressive power, we adopt real medical document collections and compare the performance of our proposed method with LDA by considering different application scenarios. The experimental results show major improvements.

The remainder of this paper is organized as follows. In the next section, we review the related work. In Section ~\ref{sec:method}, we present our fuzzy set theory based model in detail. An empirical study was conducted to verify the effectiveness of our method and the results are provided in Section ~\ref{sec:exp}. Finally, we present a summary, limitations, and future directions in Section ~\ref{sec:con}.

\section{Related Work}
\label{sec:rel}
Medical documents including clinical notes and research papers contain valuable information created by clinicians and researchers. The documents provide rich data waiting to be analyzed. The information in medical documents can be found in the form of narrative and semi-structured format. Some research have been done to extract information from medical documents \cite{friedman1994general,haug1994natural,hahn2002medsyndikate}. There are two major research area in mining medical documents. The first one tracks concepts by looking for frequency of words \cite{poulin2014predicting}. The second area categorizes the concepts to find latent variables in medical documents \cite{lin2006generative}. The first approach leads to high sparse
dimensionality data \cite{aggarwal2012introduction}; therefore, researchers have been motivated  to use the second approach such as topic modeling. Topic models can help to cluster terms representative of a particular situation such as symptoms or drugs.  The goal of topic modeling is to find common topics of dicussion in a corpus. Among topic models, LDA \cite{blei2003latent} is a popular unsupervised topic model. LDA groups words with similar semantic. Two major outputs of LDA are the probability of each topic for each document, $P(T|D)$, and the probability of each word for each topic, $P(W|T)$. This method is the most effective representation model among supervised and unsupervised topic models \cite{halpern2012comparison}.

In medical domain, LDA has been leveraged in a wide range of applications. For example, Arnold et al. (2010) used LDA for comparing the topics of patient notes \cite{arnold2010clinical} and Bisgin et al. (2011) used LDA in FDA drug side effects labels to cluster drugs \cite{bisgin2012investigating}.  Some other researches propose new variant of LDA to improve its performance for example Cohen et al. (2014) propose a topic model based on LDA to take into account the problem of redundancy in clinical notes \cite{cohen2014redundancy}.

One of methods that has not been fully considered in medical text mining is fuzzy clustering. Since Bellman and Zadeh \cite{bellman1970decision} described the decision-making method in fuzzy environments, an increasing number of studies have dealt with uncertain fuzzy problems by applying fuzzy set theory \cite{karami2012fuzzy,5609295}. Fuzzy Clustering has been used more for image analysis in medical literature \cite{saha2014multiobjective,cui2013global,beevi2012robust}. A few work have been done in medical text mining using fuzzy clustering \cite{ben2012data,fenza2012hybrid}. The main difference between our method and other document fuzzy clustering methods such as \cite{singh2011document} is that our method uses fuzzy clustering and word weighting as a pre-processing step for feature transformation before implementing any classification or clustering algorithms; however, other methods use fuzzy clustering as a final step to cluster or classify the documents. Among fuzzy clustering methods, Fuzzy C-means \cite{bezdek1981pattern} is the most popular one \cite{bataineh2011comparison}. In addition, we recently used fuzzy clustering as a feature transformation (dimension reduction) approach \cite{karami2014latent} and also as a method for topic modeling \cite{karami2015fuzzy} which uses fuzzy clustering for documents in the third step. In this research, we propose a novel method that combines local term weighting and global term weighting with fuzzy clustering to extract latent semantic features from medical documents. 
In this paper, we extract latent semantic themes from medical documents.

 \section{FLATM }
\label{sec:method}
 In this section, we detail our \textit{\underline{F}uzzy \underline{L}ogic \underline{A}pproach
 \underline{T}opic \underline{M}odel (FLATM)} and describe the steps. FLATM has seven steps with three main steps including \textit{Local Term Weighting (LTW), Global Term Weighting (GTM), and Fuzzy Clustering (FC)}. In this algorithm (Algorithm 1), the output(s) of each step is the input(s) of the next step(s). 
 
 \begin{algorithm}[ht!]
   \caption{FLATM algorithm}\label{euclid}
   \begin{algorithmic}[1]

   \Statex \textbf{Functions:}E():Entropy;I():IDF;PI():ProbIDF; NO():Normal; GFIDF:GI(); FC():Fuzzy Clustering.
  
   \State Remove stop words
   \Statex \textbf{Step 1:} Calculate LTW
   \For{\texttt{$i=1$ to $n$}}
   \For{\texttt{$j=1$ to $m$}}
   \State Calculate $f_{ij}$, $b(f_{ij})$, $p_{ij}$
  
   \EndFor
   \EndFor
  
   \Statex \textbf{Step 2:} Calculate GTW
   \For {\texttt{$i=1$ to $m$}}
   \For {\texttt{$j=1$ to $n$}}
   \State  Execute E($p_{ij}$,$n$),I($f_{ij}$,$n$),PI($b(f_{ij})$,$n$),   NO($f_{ij}$,$n$), GI($f_{ij}$,$b(f_{ij})$)
  
   \EndFor
   \EndFor

   \Statex \textbf{Step 3:} Perform Fuzzy Clustering to Find each Topic Membership for each Word $P(T_k|W_i)$
   \State Execute FC(E), FC(I), FC(PI), FC(NO), FC(GI), FC($f_{ij}$)

   \Statex \textbf{Step 4:} Calculate Each Word Probability $P(W_i)$ for each of GTW methods in step 3
   \State  $\frac{\sum_{j=1}^{n} E_{ij}}{\sum_{i=1}^{m} \sum_{j=1}^{n} E_{ij}},\frac{\sum_{j=1}^{n} I_{ij}}{\sum_{i=1}^{m} \sum_{j=1}^{n} I_{ij}},\frac{\sum_{j=1}^{n} PI_{ij}}{\sum_{i=1}^{m} \sum_{j=1}^{n} PI_{ij}},$
   \Statex $ \frac{\sum_{j=1}^{m} NO}{\sum_{i=1}^{m} \sum_{j=1}^{n} NO},\frac{\sum_{j=1}^{m} GI}{\sum_{i=1}^{m} \sum_{j=1}^{n} GI},\frac{\sum_{j=1}^{n} f_{ij}}{\sum_{i=1}^{m} \sum_{j=1}^{n} f_{ij}} $

   \Statex \textbf{Step 5:} Calculate the Joint Probability of  Word and Topic $P(W_i,T_k)$
   \State $P(T_k|W_i) \times P(W_i)$
  
   \Statex Then Calculate the Probability of  each Word in each Topic $P(W_i|T_k)$
   \State $\frac{P(W_i,T_k)}{\sum_{i=1}^{m} P(W_i,T_k)}$

   \Statex \textbf{Step 6:} Calculate the Probability of each Word in each Document $P(W_i|D_j)$
   \State $\frac{E}{\sum_{1}^{n} E}$, $\frac{I}{\sum_{1}^{n} I}$, $\frac{PI}{\sum_{1}^{n} PI}$, $\frac{NO}{\sum_{1}^{n} NO}$,
   $\frac{GI }{\sum_{1}^{n} GI}$, $\frac{f_{ij}}{\sum_{1}^{n} f_{ij}}$
  
   \Statex \textbf{Step 7:} Calculate the Probability of each Topic in each Document $P(T_k|D_i)$
   \State $\sum_{i=1}^{m} P(T_k|W_i) \times P(W_i|D_j)$
   \end{algorithmic}
   \end{algorithm}

\textbf{Step 1:} The first step is to calculate LTW. Among different LTW methods we use term frequency as a popular method. Symbol $f_{ij}$ defines the number of times term $i$ happens in document $j$. We have $n$ documents and $m$ words. Let
   \begin{equation}
   b(f_{ij})=\left\{
   \begin{array}{c l}
       1 & f_{ij}>0\\
       0 & f_{ij}=0
   \end{array}\right.
   \end{equation}

   \begin{equation}
   p_{ij}=\frac{f_{ij}}{\sum_j f_{ij}}
   \end{equation}

\noindent The outputs of this step are $b(f_{ij}),f_{ij}$ , and $p_{ij}$. We use them as inputs for the second step.

\textbf{Step 2:} The next step is to calculate GTW. We explore five GTW methods in this paper including \textit{Entropy, Inverse Document Frequency (IDF),  Probabilistic Inverse Document Frequency (ProbIDF),} \textit{Normal}, and \textit{Global Frequency Inverse Document Frequency (GFIDF)} (Table I).

  \begin{table}[h]
    \centering

    \begin{tabular}{|p{1.4cm}|p{3.5cm}|} \hline
    \textbf{Name} & \textbf{Formula} \\ \hline
    Entropy & $1+ \frac{\sum_j p_{ij}\log_2 (p_{ij})}{\log_2 n}$ \\ \hline
    IDF & $\log_2 \frac{n}{\sum_j f_{ij}}$   \\ \hline
     ProbIDF  & $\log_2 \frac{n-\sum_j b(f_{ij})}{\sum_j b(f_{ij})}$   \\ \hline
    Normal & $\frac{1}{\sqrt{\sum_j f_{ij}^2}}$ \\ \hline
    GFIDF  & $\frac{\sum_j f_{ij}}{\sum_j b(f_{ij})}$   \\ \hline

    \end{tabular}
    \caption{GTW Methods}
    \end{table}

IDF assigns higher weights to rare terms and lower weights to common terms \cite{papineni2001inverse}. ProbIDF is similar to IDF and assigns very low negative weight for the terms happen in every document \cite{kolda1998limited}. In Entropy, it gives higher weight for the terms happen less in few documents \cite{dumais1992enhancing}. Normal is used to correct discrepancies in document lengths and also normalize the document vectors. Finally, GFIDF is another scheme of IDF. By using this method words that appear once in every document or once in one document get the smallest weight. This method gives weight to words based on frequency in one document and in all documents \cite{dumais1992enhancing}. The outputs of this step are the inputs of the next step(s).

\textbf{Step 3:} Fuzzy set theory has been used to model systems that have difficulty assigning an instance to a set \cite{karami2012fuzzy}. Fuzzy clustering is a soft clustering technique that finds the degree of membership for each data point in each cluster, as opposed to assigning a data point only to one cluster. Fuzzy clustering is a synthesis between clustering and fuzzy logic. Among fuzzy clustering methods, Fuzzy C-means (FCM) \cite{bezdek1981pattern} is the most popular one \cite{bataineh2011comparison} and its goal is to minimize an objective function by considering constraints:
\begin{equation}
     Min \: \: J_q (\mu,V,X)=\sum_{k=1}^{c} \sum_{j=1}^{n} (\mu_{kj})^q DIS_{kj}^2
     \end{equation}
     subject to:
     \begin{equation}
     0 \leq \mu_{kj}\leq1;
     \end{equation}
     \begin{equation}
     \sum_{k=1}^{c} \mu_{kj}=1
     \end{equation}
     \begin{equation}
     0<\sum_{j=1}^{n} \mu_{kj} < n;
     \end{equation}
      Where:

   \begin{center}
     $n$= number of data\\
     $c$= number of clusters (topics)\\
     $\mu_{kj}$= membership value\\
     $q$= fuzzifier, $1 < q \le \infty$ \\
     $V$= cluster center vector\\
     $DIS_{kj}=d(x_j,v_k)$= distance between $x_j$ and $v_k$
   \end{center}
By optimizing eq.3:
       \begin{equation}
     \mu_{ij}= \frac{1}{\sum_{l=1}^{c} (\frac{DIS_{kj}}{DIS_{lj}})^ \frac{2}{q-1}}
     \end{equation}
      \begin{equation}
     v_i=\frac{\sum_{j=1}^{n} (\mu_{kj})^q x_j}{\sum_{j=1}^{n} (\mu_{kj})^q}
     \end{equation}
 The iterations in the clustering algorithms continue till the the maximum change in $\mu_{ij}$ becomes less than or equal to a pre-specified threshold. The computational time complexity is $O(n)$.  We use $\mu_{ij}$ as the degree of clusters' membership for each word or $P(T_k|W_i)$. Topic ($T$) is  the membership degree of a fixed vocabulary in which words with similar semantics have a higher membership degree. 
 
 One of the problems in fuzzy clustering is handling a large dimension data. In this paper, we run fuzzy clustering for 9 times with the number of clusters from 10 to 2. Each clustering step is the input for the next one. For example, the output of applying fuzzy clustering on matrix with n words (rows) and m documents (columns) with selecting 10 as the number of cluster is a matrix with n words (rows) and 10 clusters (columns) . Then we again apply fuzzy clustering on matrix with n rows and 10 columns with selecting 9 as the number of cluster. The output of these 9 steps is a matrix with n rows and 2 columns that helps us to reduce the dimension from m to 2. This matrix is the input for fuzzy clustering for the number of topics from 50 to 200 to find $P(T_k|W_i)$.
 
 \textbf{Step 4:} In this step, we use document-term matrices of step 2 with and without GTW methods to find the probability of words, $P(W_i)$, by:
 \begin{equation}
 P(W_i)=\frac{\sum_{i=1}^{n} (W_i,D_j)}{\sum_{i=1}^{m}\sum_{j=1}^{n} (W_i,D_j)}
 \end{equation}
 
 \textbf{Step 5:} The next step is to find $P(W_i|T_k)$ by first calculate:
 \begin{equation}
 P(W_i,T_k)=P(T_k|W_i) \times P(W_i)
 \end{equation}
 Then we normalize $P(W,T)$ in each topic:
 \begin{equation}
 P(W_i|T_k)=\frac{P(W_i,T_k)}{\sum_{i=1}^{m} P(W_i,T_k)}
 \end{equation}
 
 \textbf{Step 6:} We do the similar calculation in step 5 to find $P(W_i|D_j)$:
 \begin{equation}
 P(W_i|D_j)=\frac{P(W_i,D_j)}{\sum_{i=1}^{m} P(W_i,D_j)}
 \end{equation}
 
 \textbf{Step 7:} The final step is to find $P(T_k|D_j)$ by:
 \begin{equation}
 P(T_k|D_j)=\sum_{i=1}^{m} P(T_k|W_i) \times P(W_i|D_j)
 \end{equation}

\section{Experimental Results}
\label{sec:exp}
In this section, we discuss our empirical evaluation of FLATM against LDA using three measures: document classification and document modeling. In the experiment, we use Matlab packages for Chib-style estimation\footnote{\url{http://www.cs.umass.edu/~wallach/code/etm/}} and fcm\footnote{\url{http://www.mathworks.com/help/fuzzy/fcm.html}} with its default setting for implementing FCM clustering. Moreover, we use Weka\footnote{\url{http://www.cs.waikato.ac.nz/ml/weka/}} for classification evaluation, and MALLET package\footnote{\url{http://mallet.cs.umass.edu/}} with its default setting for implementing LDA.

\subsection{Datasets}
We leverage two available datasets in this research. The first dataset is a labeled corpus of English scientific medical abstracts from Springer website. It includes 41 medical journals ranging from Neurology to Radiology. In this research, we selected 5 journals including: Arthroscopy, Federal health standard sheet, The anesthetist, The surgeon, and The gynecologist with 1527 documents and 14411 terms.The second dataset called Deidentified Medical Text is an unlabeled corpus of 1607 nursing notes with 11,059 terms (Tables II\&III).

\begin{table}[!htb]
    \centering

      \begin{tabular}{|p{3cm}|p{1.5cm}|} \hline
      
         \#Documents & 1527 \\ \hline
         \#Term Tokens & 245931   \\ \hline
         \ Unique Terms & 14411  \\ \hline
         Avg Term Per Document & 96.3 \\ \hline

         \end{tabular}
          \caption{Basic Statistics for First Dataset}

     \begin{tabular}{|p{3cm}|p{1.5cm}|} \hline
     
        \#Documents & 1607 \\ \hline
        \#Term Tokens & 299449   \\ \hline
        \#Unique Term Words  & 11059  \\ \hline
        Avg Term Per Document & 124.8 \\ \hline

        \end{tabular}  
    \caption{Basic Statistics for Second Dataset}

\end{table}

\subsection{Document Classification}
The first evaluation measure is document classification on the first datasest. We use 80\% of data for training and 20\% for testing, with 5-fold cross validation. We train the models for the five classes and calculate the likelihood for the test data. We present the classification accuracy for the models. 
Table IV shows that FLATMs with Entropy, IDF, Normal, and ProbIDF have better performances than LDA in almost all different number of topics. The advantage of approach is especially obvious for a large number of topics. In addition, the combination of FLATM using GFIDF and FLATM without using any of GTW methods produces a lower performance in comparison to LDA. We remove the combination of FLATM using GFIDF and FLATM without using any of GTW methods in the rest of the experiments. Finally, the sign test shows that the improvement of FLATMs over LDA is statistically significant with a $p-value <$ 0.05.

\begin{table}[ht]
\centering

{\small
\hfill{}

\begin{tabular}{|p{1.5cm}|p{0.9cm}|p{0.9cm}|p{0.9cm}|p{0.9cm}|}
\hline

\textbf{Method}  & \textbf{50 Topics} & \textbf{100 Topics} & \textbf{150 Topics} &  \textbf{200 Topics}    \\ \hline

\textbf{FLATM (Entropy)} &       71.31 & 71.91 & 72.04& 72.88  \\ \hline

\textbf{FLATM (ProIDF)} &        70.13 & 71.71 & 71.05 & 72.76  \\ \hline

\textbf{FLATM (IDF)} &           71.57& 68.96 & 71.31 & 70.98 \\ \hline

\textbf{FLATM (Normal)}  &       69.15 & 69.81 & 70.81 & 71.57  \\ \hline

\textbf{LDA }&                   72.29 & 66.01 & 65.42 & 63.91  \\ \hline

\textbf{FLATM} &                 44.66 &  42.24 & 53.11& 49.91  \\ \hline

\textbf{FLATM (GFIDF)} &         38.51 & 38.70 & 38.83& 39.03  \\ \hline

\end{tabular}}
\caption{Document Classification Accuracy (\%)}
\end{table}

 \begin{figure*}[ht]
  \centering
      \begin{subfigure}[ht]{0.35\textwidth}
          \centering
          \resizebox{\linewidth}{!}{
              \begin{tikzpicture}[scale=.6]
   \begin{axis}[ xlabel={Number of Topics}, ylabel={Log-Likelihood},legend style={at={(0.5,-0.2)}, anchor=north,legend columns=-1}]
    \addplot coordinates { (25,-104843.460603056)(50,-104843.460603056)(75,-104843.460603057) (100,-104842.235505143)(125,-104842.74116663) (150,-104842.123930449)(175,-104842.629797738) (200,-104843.460603057) };
     \addplot coordinates { (25,-105014.627139032)(50,-105016.779448629)(75,-105015.578608512) (100,-105015.639350093)(125,-105017.076759624) (150,-105016.265313158)(175,-105017.122662854) (200,-105017.076454177) };
     \addplot coordinates { (25,-110226.96571032)(50,-110226.96571032)(75,-110227.782091809) (100,-110229.612066465)(125,-110225.856190989) (150,-110230.047625255)(175,-110228.4252726) (200,-110230.23325156)};
      \addplot coordinates { (25,-106122.156600721)(50,-106123.736795348)(75,-106124.504717432) (100,-106124.087343569)(125,-106124.085645785) (150,-106125.222767848)(175,-106124.616250361) (200,-106125.231267389) };
     \addplot coordinates { (25,-121805.811731416)(50,-121683.091659729)(75,-121805.811731416) (100,-121686.572984534)(125,-121899.021708174) (150,-121730.512963274)(175,-121845.380117282) (200,-121467.948683721) };
       \end{axis}
  
  \end{tikzpicture}
          }
          \caption{First Dataset}
          \label{fig:subfig8}
      \end{subfigure}
      \begin{subfigure}[ht]{0.35\textwidth}
      \centering
          \resizebox{\linewidth}{!}{
              \begin{tikzpicture}[scale=.6]
  \begin{axis}[ xlabel={Number of Topics}, ylabel={Log-Likelihood},legend style={at={(0.5,-0.2)}, anchor=north,legend columns=-1}]
   \addplot coordinates { (25,-137867.865372641)(50,-137867.851049937)(75,-137869.340619458) (100,-137869.621793458)(125,-137868.894279709) (150,-137868.406132936)(175,-137869.829915297) (200,-137869.86589945) };
    \addplot coordinates { (25,-138606.277287529)(50,-138606.706397826)(75,-138606.993235349) (100,-138606.545848352)(125,-138607.245957813) (150,-138609.334232713)(175,-138608.538605372) (200,-138608.370353161) };
    \addplot coordinates { (25,-145790.640922964)(50,-145791.387504195)(75,-145790.840169173) (100,-145792.240426892)(125,-145790.693736896) (150,-145792.036927813)(175,-145791.058719099) (200,-145793.036808923) };
    \addplot coordinates { (25,-138352.16)(50,-138352.164432488)(75,-138488.057669703) (100,-138586.608340486)(125,-138726.22625622) (150,-138832.823302625)(175,-138944.193220478) (200,-138944.193220478) };
    \addplot coordinates { (25,-179351.930737117)(50,-178786.751372205)(75,-179006.319810626) (100,-176510.592554451)(125,-178645.206656811) (150,-177969.385829714)(175,-177600.481432976) (200,-177119.616552266) };
      \end{axis}
  
  \end{tikzpicture}
          }
          \caption{Second Dataset}   
          \label{fig:subfig9}
      \end{subfigure}
     

        

  
  

  \begin{tikzpicture}

      \begin{customlegend}[legend columns=-1,
        legend style={
          draw=none,
          column sep=1ex,font=\tiny
        },legend entries={$FLATM(Entropy)$,$FLATM(IDF)$,$FLATM(ProbIDF)$,$FLATM(Normal)$,$LDA$}]
        
      \addlegendimage{blue,fill=blue,mark=*,sharp plot}
      \addlegendimage{red,fill=red!50!black,mark=square*,sharp plot}
      \addlegendimage{brown,fill=black!50!brown,mark=otimes*,sharp plot}
      \addlegendimage{black,fill=black,mark=star,sharp plot}
      \addlegendimage{blue,fill=blue,mark=diamond*,sharp plot}
      \end{customlegend}
  \end{tikzpicture}
  
 \caption{Likelihood Comparison} 
  \label{fig:subfig1.a.4}
  \end{figure*}
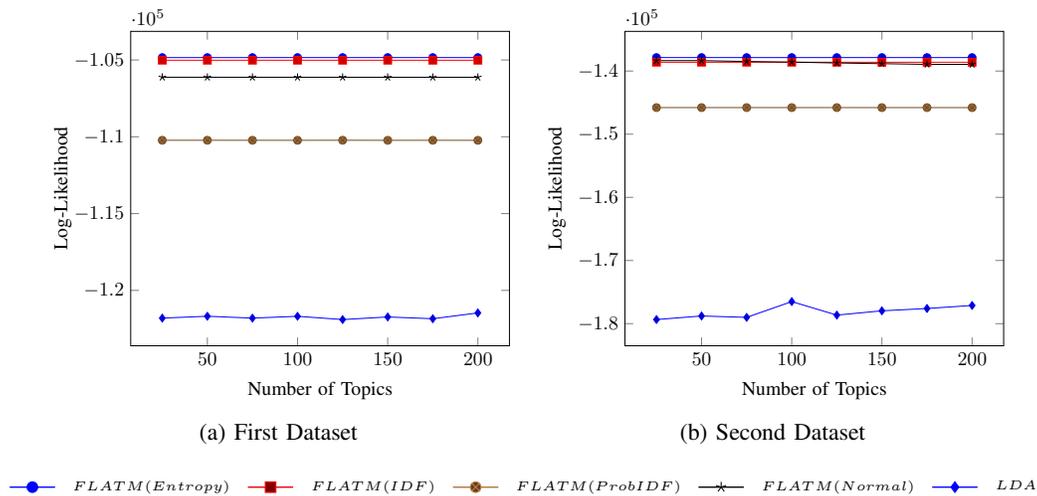

 \subsection{Document Modeling}
 The third evaluation measurement is document modeling using log-likelihood. We trained FLATMs and LDA, on both datasets to compare generalization performance of these models. The documents in the corpora are treated as unlabeled; thus, our goal is density estimation to achieve high likelihood on a held-out test set.  
 We  split the first and the second dataset into two sublets with 90\% and 10\% of the dataset respectively. In preprocessing the data, we removed a standard list of stop words from each corpus. Then we learn topics from the larger set and calculate log-likelihood for the smaller set, $P(D_{test}|T)$. 
 
 There are different methods to calculate log-likelihood that among them Chib-style estimation shows better performance \cite{wallach2009evaluation}. In this evaluation part, we remove FLATM with GFIDF and without any GTW because of their week performance in document classification, and we focus on LDA and the rest of FLATMs. We compare FLATMs with LDA and the result shows that FLATMs have a better performance over LDA with different number of topics and different sets of training data.
 
 \newpage

  Figures 2.a and 2.b present the log-likelihood for each model on both corpora for different number of topics. FLATMs consistently perform better than LDA.

\section{Conclusion}
\label{sec:con}
A large volume of medical data has been accumulated in recent years. Analyzing such data is becoming more and more important to advance state-of-the-art healthcare. Due to the unstructured nature of free-text format for the medical data, text mining techniques such as topic modeling are widely adopted to extract latent semantic
properties of a medical corpus. 

Despite the usefulness of topic models for medical data analysis, existing topic models such as LDA still suffer from several critical issues, such as extremely high computational complexity and unsatisfactory performance for data analytical tasks. In this paper, we proposed the use of fuzzy set theory, the fuzzy clustering technique in particular, for modeling unstructured medical documents. Fuzzy clustering is one of the machine learning techniques that has been used more in medical image processing. To the best of our knowledge, this is the first study that uses fuzzy clustering for topic modeling of medical documents. Our proposed method, FLATM, is a topic model that uses fuzzy clustering with local and global term weighting methods to disclose the latent semantic of medical documents.

Compared to LDA, FLATM has a much lower computational complexity and provides stronger expressive power of medical documents. Our empirical evaluation conducted on several real medical datasets showed that FLATM outperforms LDA in various data analytical tasks including medical document classification and modeling. 

There are several interesting directions to explore in the future. For example, prediction of stages of various diseases is very important in healthcare. We are interested in applying our FLATM method on large-scale medical data to provide accurate predictions for patients like using in home healthcare robots \cite{alaiad2014determinants1,alaiad2014exploratory,alaiad2013patients}. In addition, our approach can be used for other domains such as spam detection in SMS and online reviews \cite{karami2014asist,karami2015online,DBLP:conf/amcis/KaramiZ14}.


\bibliographystyle{IEEEtran}
\bibliography{acl2014}


\end{document}